\documentclass[twocolumn,english]{revtex4-1}

\usepackage[T1]{fontenc}
\usepackage{amsbsy}
\usepackage{amsmath}
\usepackage{amstext}
\usepackage{amssymb}
\usepackage{graphicx}
\usepackage{units}
\usepackage{babel}

\begin{document}

\title{Proposal for a tunable graphene-based terahertz Landau-level laser}

\author{Samuel Brem}
\email{samuel.brem@physik.tu-berlin.de}
\author{Florian Wendler}
\affiliation{Technical University Berlin, Institute of Theoretical Physics, 10623 Berlin, Germany}
\author{Ermin Malic}
\affiliation{Chalmers University of Technology, Department of Applied Physics, 41296 Gothenburg, Sweden}

\begin{abstract}
In the presence of strong magnetic fields the electronic bandstructure of
graphene drastically changes. The Dirac cone collapses into discrete non-equidistant
Landau levels, which can be externally tuned by changing the magnetic field. In contrast 
to conventional materials, specific Landau levels are selectively addressable using 
circularly polarized light. Exploiting these unique properties, we propose the design of a tunable laser operating in the technologically promising terahertz spectral range. To uncover the many-particle physics behind the emission of light, we perform a fully quantum mechanical investigation of the non-equilibrium dynamics of electrons, phonons, and photons in optically pumped Landau-quantized graphene embedded into an optical cavity. The gained microscopic insights allow us to predict optimal experimental conditions to realize a technologically promising terahertz laser.
\end{abstract}
\maketitle

The terahertz (THz) regime of the electromagnetic spectrum can be exploited in a wide range of applications including medical diagnostics, atmosphere and space science as well as security and information technology 
\cite{tonouchi2007cutting, Pickwell2006biomedical, liu2007terahertz, Federici2010thzComm}. 
Although THz research has progressed significantly in the last 20 years, the transition from laboratory demonstration to practical environment has occurred slowly and only for some niche applications.
The largest challenge is the lack of adequate, tunable THz radiation sources. In 1986, H. Aoki proposed to design Landau level (LL) lasers exploiting the discreteness of LLs in two-dimensional electron gases \cite{Aoki1986novelLLL}. Here, the energetic LL spacing and thus the possible laser frequency can be externally tuned through the magnetic field. The challenge for the realization of such a laser is to obtain a stable population inversion, i.e. a larger carrier occupation within an energetically higher LL. Since conventional semiconductors exhibit an equidistant spectrum of LLs, strong Coulomb scattering acts in favor of an equilibrium Fermi-Dirac distribution and strongly counteracts the build-up of a population inversion. In contrast, graphene as a two-dimensional material with a linear dispersion exhibits a non-equidistant LL separation offering entirely different conditions for many-particle processes \cite{Aoki2009Cyclotron,wendler2015towards,Wang2015Continuous}.  
Exploiting these remarkable properties of Landau-quantized graphene, we propose an experimentally accessible scenario to achieve continuous wave lasing with tunable frequencies in the technologically promising terahertz spectral regime.

\begin{figure}[!t]
\begin{centering}
\includegraphics[width=0.75\columnwidth]{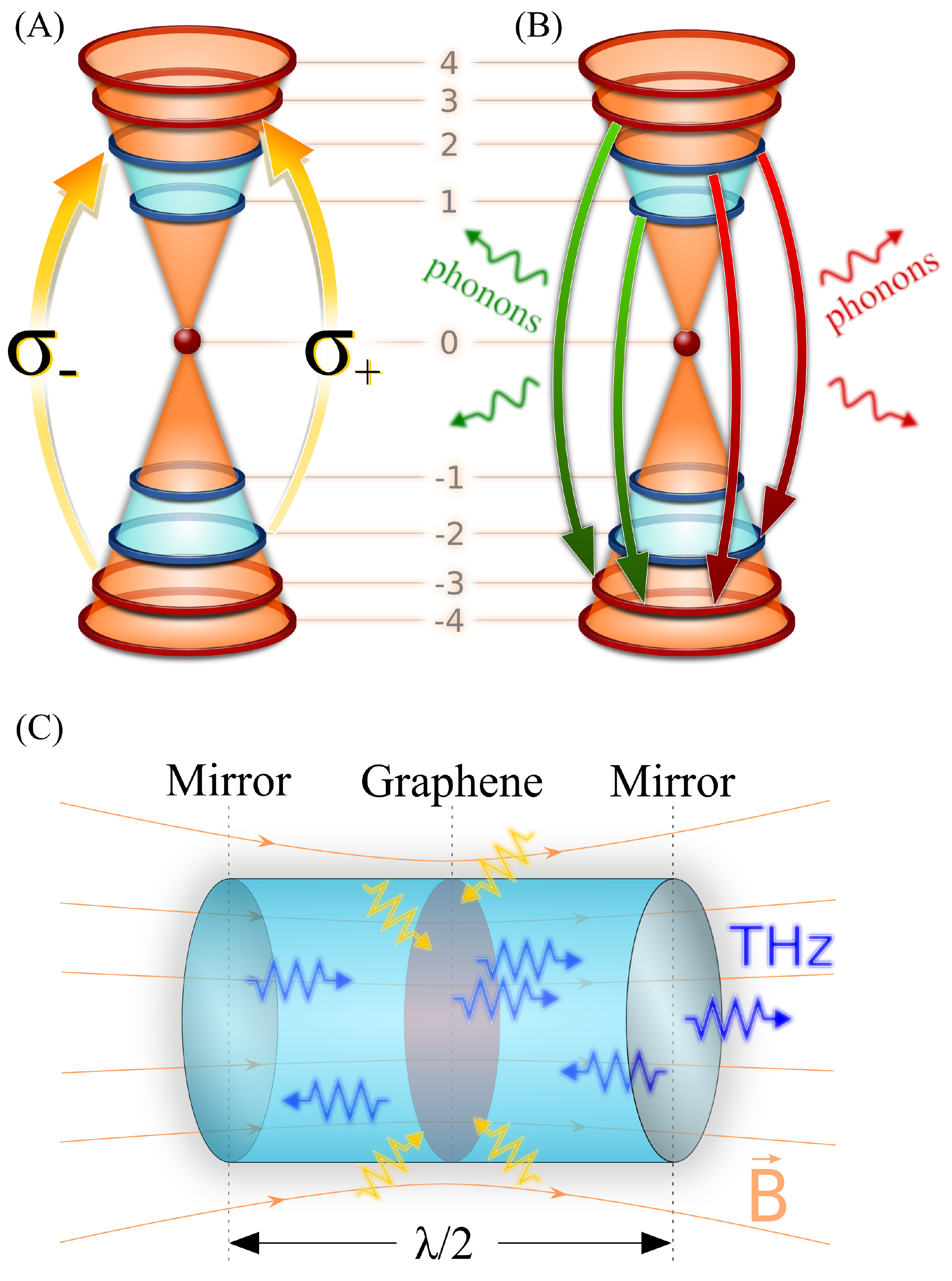} 
\end{centering}
\caption{\textbf{Laser scheme.} 
Sketch of the energetically lowest Landau levels in graphene embedded in an optical cavity. (A) The linearly polarized optical pump field
 induces transitions  $l=-3 \rightarrow +2$ and $l=-2 \rightarrow +3$ (yellow arrows), which results in a population inversion
between $l=+1$ and $l=+2$ in the conduction band and $l=-1$ and $l=-2$ in the valance band. (B) The emission of optical phonons can act in favor of the laser cycle (green arrows) or counteract the build-up of the population inversion (red arrows). (C) To achieve stimulated emission of photons, the system is embedded into an optical cavity. }
\label{fig:scheme} 
\end{figure}

\noindent The non-equidistant arrangement of energy levels\cite{Goerbig2011Electronic,Sadowski2006Spectroscopy}
\begin{equation}\label{eq:epsLL}
 \varepsilon_l = sgn\{l\}\hbar v_F \sqrt{\dfrac{2 e_0 B}{\hbar} \lvert l \rvert},
\end{equation}
combined with selection rules for circularly polarized light, allow to selectively address individual inter-LL transitions. Here,  the magnetic field $B$ is perpendicular to the graphene layer, $v_F$ denotes the Fermi velocity 
in graphene, and $l=...,-2,-1,0,1,2, ...$ is the LL quantum number. Left (right) circularly
polarized light, denoted with $\sigma_{+(-)}$, exclusively induces transitions with quantum
numbers \cite{rao2012coherent, wendler2014carrier} $\lvert l\rvert \rightarrow \lvert l  \rvert +(-) 1$ .  Thus, a linearly polarized optical pump field with
an energy matching the transition $l=-2 \rightarrow +3$ can simultaneously induce a population inversion 
between $l=+2$ and $l=+1$ as well as $l=-1$ and $l=-2$, cf. Fig. \ref{fig:scheme}A. In contrast to conventional materials, the 
non-equidistant spectrum of LLs in graphene efficiently quenches Coulomb scattering due to restrictions stemming from the energy conservation. 

While in a previous study \cite{wendler2015towards}, we have predicted the appearance of such a population inversion
 between optically coupled LLs, in this work we go a significant step forward. To achieve laser light emission, we propose to embed the graphene layer into a high quality Fabry-Perot microcavity \cite{engel2012light} with a resonator mode matching the energy difference between  $l=+1$ and $l=+2$, cf. Fig. \ref{fig:scheme}C. 
This way, the trapped cavity photons become multiplied via stimulated emission, generating coherent 
terahertz radiation. The lasing process including optical pumping and stimulated emission of photons is complemented by the emission of 
optical phonons. The latter depopulate the lower laser level $l=+1$  and repopulate the initial LL $l=-3$  for optical excitation, cf. inner green arrow in Fig. \ref{fig:scheme}B. As a result, carriers perform cycles in a three-level system and continuous wave lasing is possible. 
To model the laser dynamics, we develop a fully quantum mechanical theoretical approach providing microscopic access to the coupled electron, phonon, and photon dynamics of Landau-quantized graphene. The gained insights allow to predict optimal experimentally accessible conditions including magnetic field, cavity quality factor, and pump intensity. 

\section*{Microscopic Modell}
Based on the density matrix formalism in second quantization \cite{haug1990quantum,Rossi2002Theory,kira2006many, malic2013graphene} combined with tight binding wave functions \cite{Reich2002Tight,Goerbig2011Electronic,wendler2015ultrafast}, we derive
a set of luminescence Bloch equations:

\begin{eqnarray}
\dot{\rho}_{l}(t) & = & 2 \sum_{\mu l'} \Re\{\lvert g^{\mu}_{l'l}\rvert^2 S^{\mu}_{ll'} - \lvert g^{\mu}_{ll'}\rvert^2 S^{\mu}_{l'l}\} \\
\nonumber & & + \sum_{l'} P_{ll'}\big(\rho_{l'}-\rho_{l}\big) +  \Gamma_l^{\text{in}}(t)\big(1-\rho_{l}\big) -\Gamma_l^{\text{out}}(t)\rho_{l},  \label{eq:rhopunkt}\\
\dot{n}_{\mu}(t) & = & 8 N_B  \sum_{ll'} \lvert g^{\mu}_{ll'}\rvert^2 \Re \{ S^{\mu}_{l'l}\} -2\kappa \big(n_{\mu}-n_0\big),\label{eq:npunkt} \\
\dot{S}^{\mu}_{ll'}(t) & = & i\big(\omega_{ll'}+\omega_{\mu}+i\gamma_{ll'}(t)+i\kappa\big)S^{\mu}_{ll'} \nonumber\\
& &+  n_{\mu}\big(\rho_{l'}-\rho_{l}\big) +\rho_{l'}\big(1-\rho_{l}\big) +  T^{\mu}_{l'} - T^{\mu}_{l}  \label{eq:Spunkt} .
\end{eqnarray}
This system of coupled differential equations describes the temporal evolution of the carrier occupation probabilities
$\rho_l(t)$ of LLs with the quantum number $l$. The carrier occupation is coupled to the number of photons $n_{\mu}(t)$ of the two relevant cavity modes $\mu=\sigma\pm$ via the
photon-assisted polarization $S^{\mu}_{ll'}(t)$, which is the probability amplitude for emitting a $\mu$-photon via transitions $l'\longrightarrow l$.

The process of optical pumping enters the equations through the pump rate $P_{ll'}$. The carrier-carrier and carrier-phonon 
interactions are treated within a correlation-expansion on a two particle level\cite{Malic2011Microscopic}, which leads to time- and energy-dependent in- and 
out-scattering rates $\Gamma_l^{\text{in/out}}(t)$. Those incorporate all electron-electron and electron-phonon
scattering channels, including time-dependent Pauli blocking terms. The Coulomb interaction is calculated by taking into account many-particle and static dielectric screening induced by the substrate \cite{Goerbig2011Electronic,wendler2014carrier}. Within the relevant momentum regime, the energies of acoustic phonons are too small to induce inter-Landau level transitions.
Therefore, carrier-phonon scattering is only considered for the dominant optical phonon modes ($\Gamma \text{TO}$, $\Gamma \text{LO}$, $\text{KTO}$ and $\text{KLO}$) \cite{piscanec2004kohn,Maultzsch2004Phonon} in a bath approximation \cite{wendler2015ultrafast}. The energy conservation of all interactions is softened due to the finite dephasing of coherences resulting in a  broadening of LLs. The dephasing rates $\gamma_{ll'}$ are determined  self-consistently considering 
many-particle and impurity-induced scattering \cite{ando1974theory,shon1998quantum,funk2015microscopic}. Details of the calculation including expressions for the pump and scattering rates, as well as the self-consistent determination of the dephasing rates can be found in the supplementary material.

The interaction strength between electrons and cavity photons is determined by the coupling 
element $g^{\mu}_{ll'}$. Furthermore, the photon generation rate is influenced by the number of emitters that is given by the magnetic field dependent LL degeneracy $N_B=BA/\Phi_0$ corresponding to the number of magnetic flux quanta $\Phi_0=h/e_0$ within the graphene layer of area $A$. Since the electron-photon coupling
is relatively weak in graphene, spontaneous emission into non-lasing modes is negligibly small. We
consider a finite cavity photon lifetime $(2\kappa)^{-1}= Q / \omega_{\sigma\pm}$ that is given by the quality factor $Q$ and the photon frequency
$\omega_{\sigma+}=\omega_{\sigma-}$. Therewith we account for photon losses due to cavity imperfections and laser light out-coupling, which leads to a decay of the photon number towards a thermal occupation $n_0$. 
To prove whether coherent laser light is emitted from graphene, we also track the temporal evolution of the photon statistics via the second-order correlation
function  $g^{(2)}(t,\tau)$. For zero delay time $\tau$, it is a measure for the quantum mechanical intensity fluctuations of the emitted light\cite{scully1997quantum}.
Coherent laser light is characterized by $g^{(2)}(t,0)=1$, whereas $g^{(2)}(t,0)>1$ holds for thermal and $g^{(2)}(t,0)<1$ for non-classical light. 
To calculate $g^{(2)}$ we consider the evolution of photon-photon and higher electron-photon-correlations (as for example $T^{\mu}_{l}$ in Eq. \ref{eq:Spunkt}) up to the quadruplet level\cite{Gies2007Semiconductor,Jago2015Graphene}, cf. supplementary material.

\section*{Results}

The solution of the luminescence Bloch equations reveals the non-equilibrium dynamics of the electronic configuration and the number of photons within the cavity, which provides a microscopic understanding of the switch-on characteristics of the Landau level laser. In the following, we investigate the dynamics at room temperature and the following experimentally accessible conditions: the cavity cross-section area  $A=\unit[100]{\mu m^2}$ (also size of the graphene sheet, cf. Fig. \ref{fig:scheme}C), the cavity length is fixed due to the resonance condition $L=\lambda_\mu/2$, a quality factor \cite{vahala2003optical,chen2014terahertz} of $Q=\unit[5000]{}$, a background screening $\varepsilon_{bg}=3.3$ (corresponding to a SiC substrate), and a reasonable impurity strength \cite{mittendorff2015carrier} determined by an impurity-induced LL broadening of $\unit[2.5] {meV}$ at $B=\unit[4]{T}$.

\begin{figure}[!t]
\begin{centering}
\includegraphics[width=0.85\columnwidth]{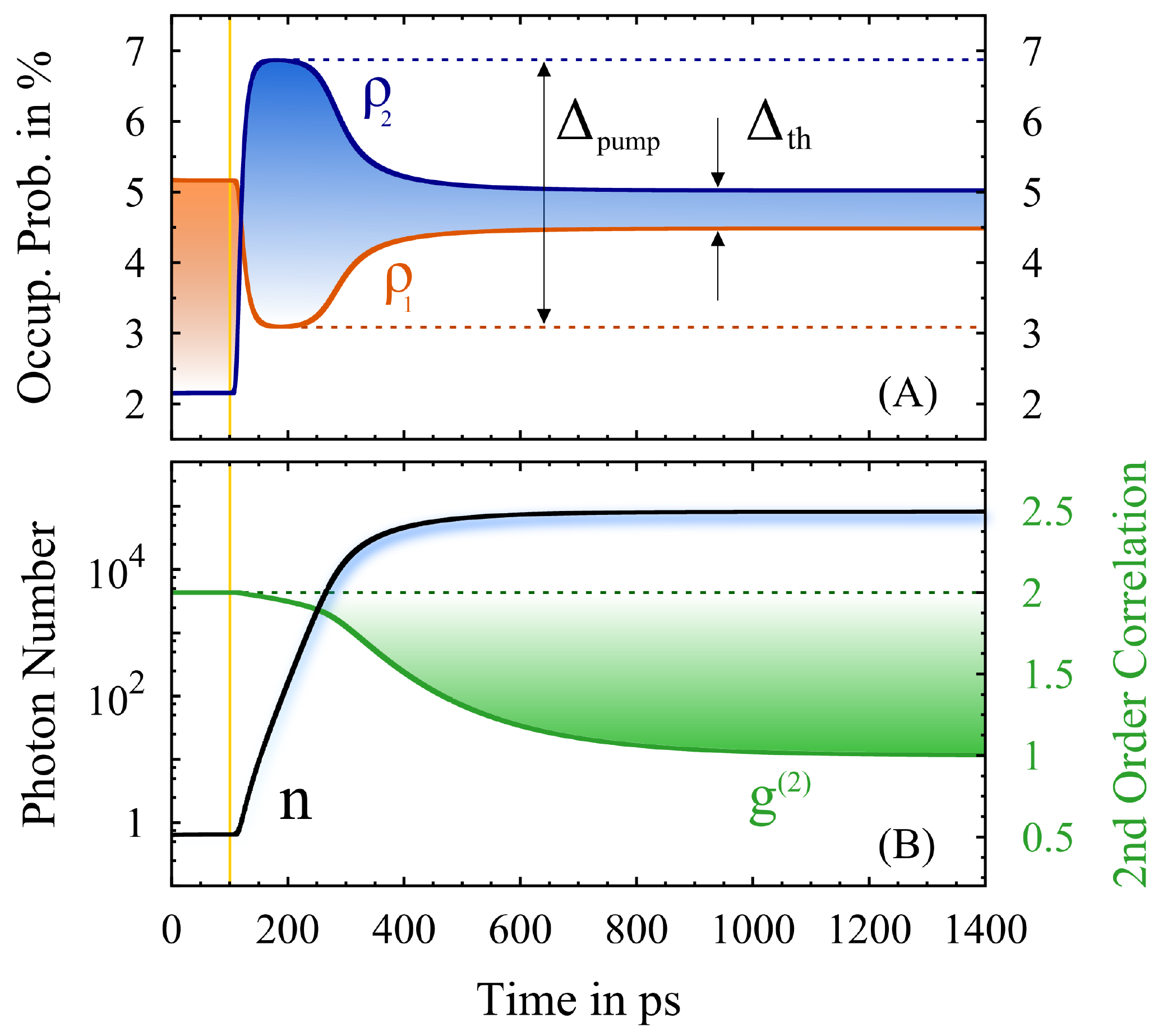} 
\par\end{centering}
\caption{\textbf{Laser dynamics.} 
(A) Time development of the occupation probabilities of the two laser levels  in the conduction band $\rho_2$ and $\rho_1$ at $B=\unit[4]{T}$. 
The thermal electron population at $t=0$ is inverted (blue shaded area) after the pump field is turned on (yellow line).
Without stimulated emission, the inversion would stay at the indicated value $\Delta_\text{pump}$ (dashed lines).
(B) Evolution of the right circularly polarized photon number  $n_{\sigma-}$ (logarithmic) and the second order correlation 
$g^{(2)}_{\sigma-}$ (right axis). The population inversion induces an exponential increase of the photon number by stimulated emission.
Due to the finite pump and relaxation rates, the population inversion is depleted with rising photon number, resulting in a quasi-stationary threshold
inversion $\Delta_{\rm{th}}$. During the stable laser equilibrium the system emits coherent laser light characterized by the second-order correlation function $g^{(2)}_{\sigma-}=1$.
}
\label{fig:transient} 
\end{figure}

\textbf{Laser dynamics.} 
At first we study the laser dynamics at the fixed magnetic field $B=\unit[4]{T}$ and the pump intensity $I=\unit[10]{kW/cm^2}$. Since in undoped graphene, the electron 
and hole populations within conduction and valence band are fully symmetric, we focus on the electron dynamics in the following. Figure \ref{fig:transient}A shows the temporal evolution of the electron occupation probability of the two laser levels  $l=+2$ and l=$+1$.  Initially both occupations are in thermal equilibrium characterized by a Fermi-Dirac distribution with $\rho_{1}>\rho_{2}$. At $\unit[100]{ps}$, the constant optical pump field is turned on transferring carriers from $l=-3$ to $l=+2$ giving rise to a population inversion with $\rho_{2}>\rho_{1}$, cf. the blue-shaded region in Fig. \ref{fig:scheme}A.

Phonon-induced relaxation of excited carriers counteracts the optical excitation mainly through transitions $l=+2 \rightarrow -2$ and $l=+2 \rightarrow -3$, cf. red arrows in Fig. \ref{fig:scheme}B. Coulomb-induced scattering is strongly suppressed due to the non-equidistant nature of the optically excited Landau levels. Besides the increase of $\rho_2$, the pump process indirectly leads to a  decrease of $\rho_{1}$, since $\rho_{-3}$ is optically depopulated opening up the channel for phonon scattering via $l=+1 \rightarrow -3$, which is sketched as the inner green arrow in Fig. \ref{fig:scheme}B. 
Shortly after the pulse is switched on, a  quasi-equilibrium between optical excitation and relaxation due to the emission of phonons is reached resulting in the pump-induced population inversion $\Delta_{\text{pump}}$.

Including an optical cavity, the number of photons increases exponentially via stimulated emission, once a population inversion is established. Figure \ref{fig:transient}B shows the time evolution of the right circularly polarized photon number $n_{\sigma+}$. The chain reaction of stimulated emissions requires more than $\unit[100]{ps}$ to generate a significant number of photons reflecting the weak electron-light interaction in graphene and the finite cavity photon lifetime.

The growing photon avalanche is accompanied by a decrease of the population inversion, due to the finite pump and relaxation rates. The occupation of the upper laser level $\rho_2$
decreases, since the stimulated emission of photons via $l=+2 \rightarrow +1$ breaks the balance between pumping and phonon relaxation. Similarly, the finite lifetime of $l=+1$ leads to an accumulation
of carriers resulting in an enhanced $\rho_1$, cf. Fig. \ref{fig:transient}A. After approximately $\unit[500]{ps}$, a new quasi-equilibrium is reached that is characterized by a reduced threshold population inversion $\Delta_{\rm{th}}$. At that point, gain and cavity losses compensate each other and the number of photons remains constant. 
Hence, to enter the laser regime, the pumped inversion $\Delta_{\text{pump}}$ has to be larger than $\Delta_{\rm{th}}$. 
An analytic expression for the threshold population inversion can be extracted from the static 
limit of \eqref{eq:npunkt} and \eqref{eq:Spunkt} by only considering the resonant polarization $S^{\sigma+}_{12}$ and
by neglecting spontaneous emission $\propto \rho_{2}(1-\rho_{1})$ and higher-order photon correlations yielding
\begin{equation} \label{eq:/Delta_th}
\Delta_{\rm{th}}=\dfrac{\kappa(\kappa+\gamma_{12})}{4 N_B \lvert g_{12}^{\sigma+}\rvert^2}.
\end{equation} 
The larger the cavity losses ($\propto \kappa$) and the faster the decay of the polarizations ($\propto (\kappa+ \gamma_{12})$), the higher the threshold inversion. On the other hand, a large number of emitters ($\propto N_B$) and a strong carrier-photon coupling $g_{12}^{\sigma+}$ result in a higher photon generation rate and therefore act in favor of a low threshold.

Finally, to describe the statistics of the emitted photons, we calculate the second-order correlation function $g^{(2)}$, cf. 
the right y-axis in Fig. \ref{fig:transient}B. Initially before the optical pumping, we find $g^{(2)}=2$ characterizing photons in thermal equilibrium. Once a population inversion is reached, the number of photons increases due to stimulated emissions, and $g^{(2)}$ approaches the value 1 characterizing coherent laser light. 

\begin{figure}[!t]
\begin{centering}
\includegraphics[width=0.85\columnwidth]{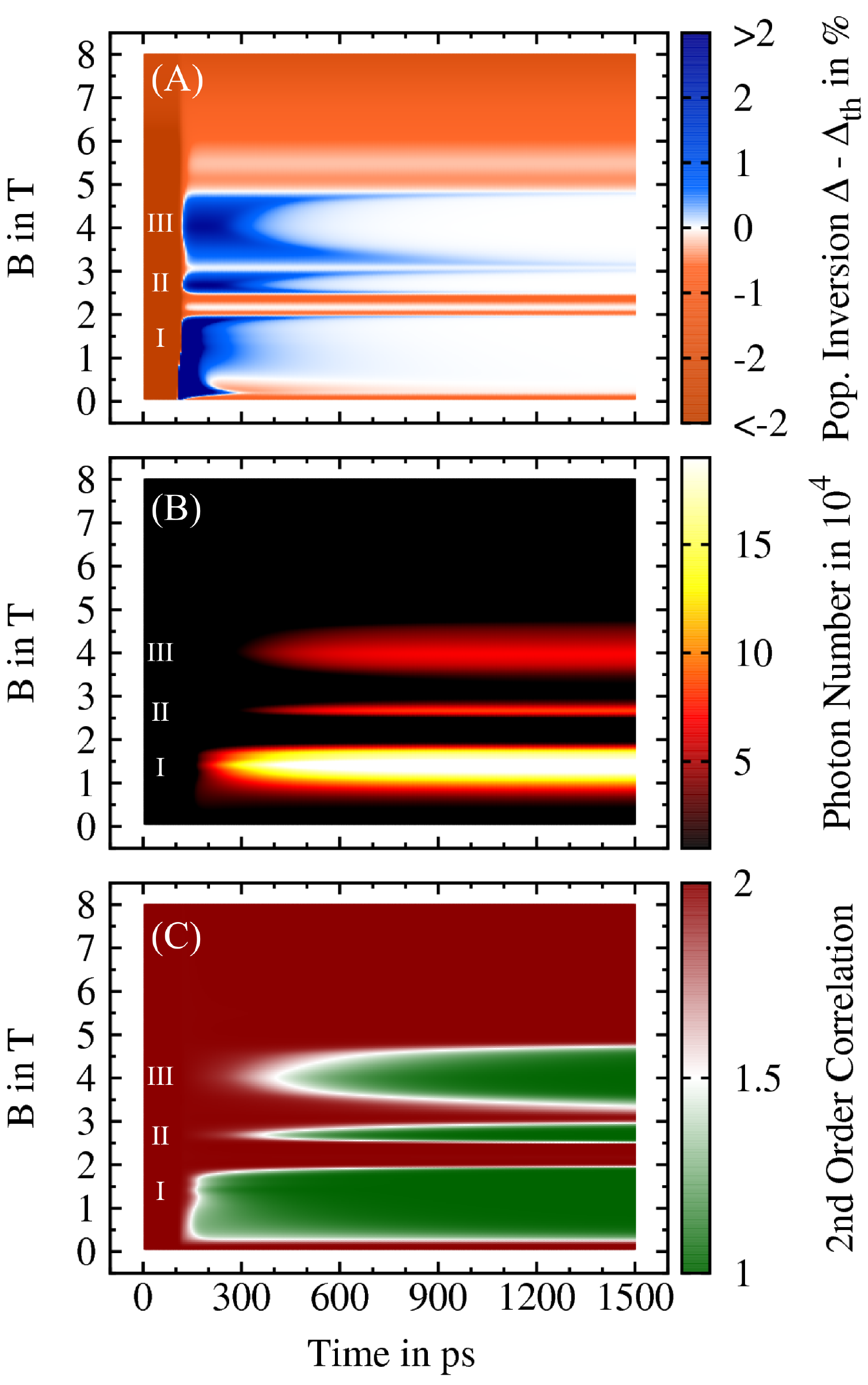} 
\par\end{centering}
\caption{\textbf{Laser tunability.} 
Time development of the (A) population inversion, (B) photon number, and (C) the second order correlation function for technologically relevant magnetic fields at constant pump intensity of
$I=\unit[10]{kW/cm^2}$.
The multiple number of optical phonon modes, which can either support or counteract the laser operation lead to a remarkable magnetic field dependence of the laser dynamics. There are
three regimes (marked with I, II, and III), where the population inversion is large enough to generate a significant number of photons. Within these magnetic field regimes the system produces coherent
THz radiation characterized by $g^{(2)}=1$.}
\label{fig:tune} 
\end{figure}

\textbf{Tunability of the laser frequency.} 
A crucial advantage, of the LL laser is its tunabilty, since the spacing between LLs is adjustable through the magnetic field. However, to allow carriers to perform cycles within the three-level laser system, a non-radiative decay $l=1\longrightarrow-3$ (and $l=3\longrightarrow-1$) via the emission of optical phonons is required, which have discrete energies.
The non-trivial interplay of the multiple phonon modes gives rise to a very interesting magnetic field dependence of the laser 
dynamics. Figure \ref{fig:tune} shows the temporal evolution of (A) the population inversion, (B) the photon number $n_\mu$, and (C) second-order correlation 
function $g^{(2)}$ for the technologically relevant magnetic fields $B$ at a constant pump intensity of $I=\unit[10]{kW/cm^2}$. 

The length $L$ of the cavity is adjusted to the $B$-dependent resonance condition $L=\lambda_\mu/2=\pi c \hbar / (\varepsilon_2-\varepsilon_1)$. Moreover, the pump frequency is changed to match the transition $l=-3\longrightarrow 2$. Since, the distance between LLs and also their broadening increases with the magnetic field, the pump rate $P \propto I/(\omega^2\gamma)$ (cf. supplementary material) decreases.
Thus, at magnetic fields $B> \unit[5]{T}$ the pump intensity is not sufficient to exceed the threshold population inversion $\Delta_{\rm{th}}$.  At very low magnetic fields $B<\unit[0.5]{T}$, the separation of LLs becomes too small to selectively pump a single LL transition, so that neighboring LLs are also pumped. As a result, the population inversion completely vanishes within the first 200ps, cf. Fig. \ref{fig:tune}(A). Between $\unit[0.5]{T}$ and $\unit[5]{T}$, we find  three distinguished zones (marked with I, II, and III), where lasing takes place. The thermal equilibrium at $t=0$ is pumped to an intermediate quasi-equilibrium at $t=\unit[100]{ps}$. Only if the achieved population inversion $\Delta_{\text{pump}}$ significantly exceeds $\Delta_{\rm{th}}$ (blue areas in Fig. \ref{fig:tune}A), stimulated emission can induce a photon avalanche and the photon number exponentially increases, cf. yellow areas in Fig. \ref{fig:tune}B. The time scale of that process strongly depends on how large $\Delta_{\text{pump}}$ is and how well phonon-induced processes assist the laser cycle. The green zones in 
Fig. \ref{fig:tune}C further illustrate that the three regimes of Fig. \ref{fig:tune}A coincide with the emission of coherent laser light. 

\begin{figure}[!t]
\begin{centering}
\includegraphics[width=0.85\columnwidth]{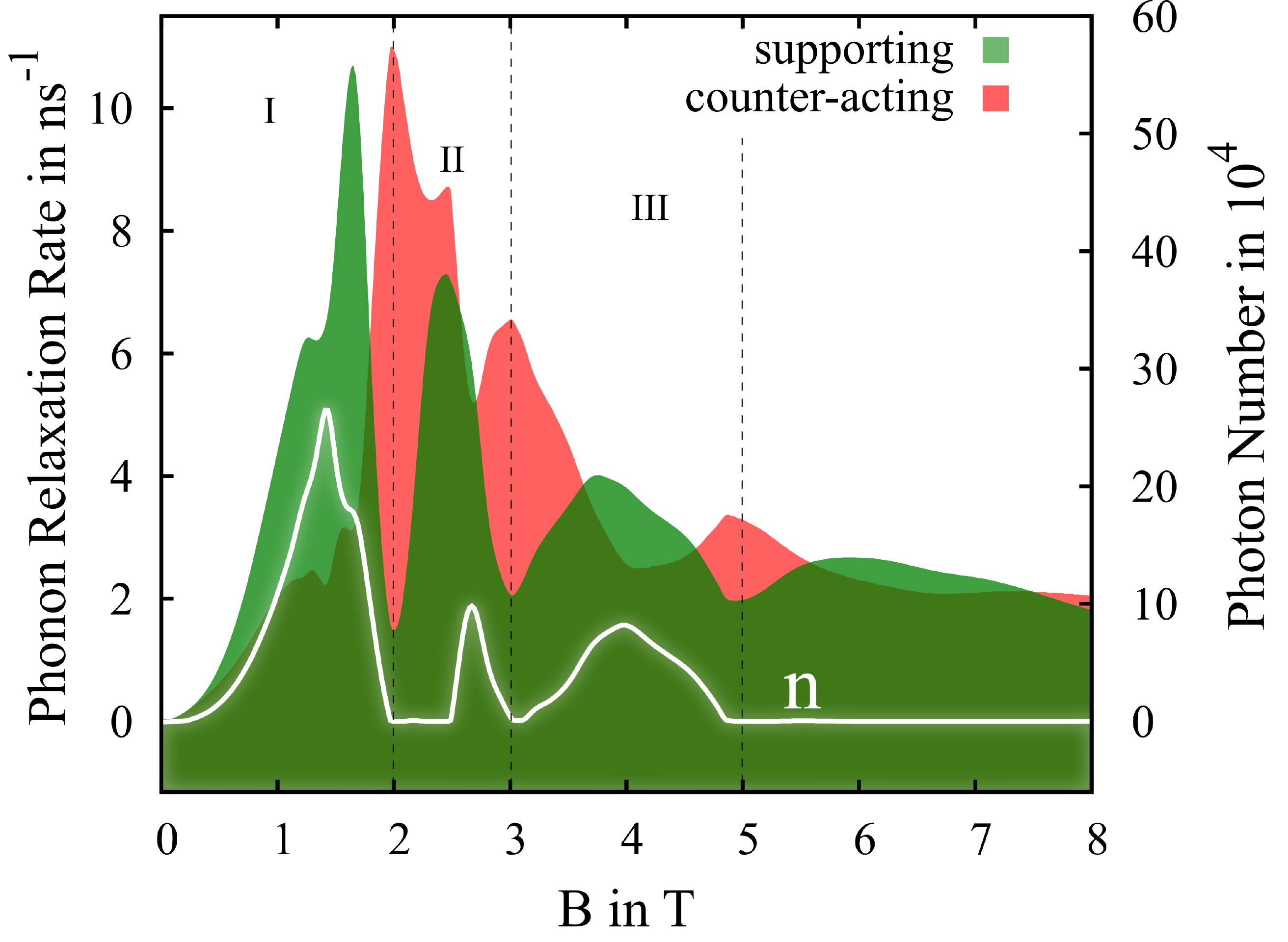} 
\par\end{centering}
\caption{\textbf{Interplay of phonon resonances.} 
Magnetic field dependence of the net phonon relaxation rates during the quasi-equilibrium, classified as laser cycle supporting (green filled curve) and counteracting channels (red filled curve). 
Photons (white curve) are only generated in regimes where supporting processes are resonant. Phonon-assisted counteracting channels deplete the population inversion 
and thus damp the generation of photons.}
\label{fig:nonrad} 
\end{figure}

The appearance of the three distinct magnetic field zones can be well understood by examining the $B$-dependence of the phonon relaxation rates. In particular, we distinguish between laser supporting channels and processes which deplete the population inversion, cf. green and red arrows in Fig. \ref{fig:scheme}B. The corresponding net phonon relaxation rates (sum over all phonon modes) during the quasi-equilibrium are illustrated in Fig. \ref{fig:nonrad}.
A significant number of photons (white curve) is only generated in $B$ regimes, where the laser cycle is effectively supported by phonon-assisted non-radiative processes (green filled curve).

In regime I, K-phonons are resonant with the transition $l=+3 \rightarrow -3$ (outer green arrow in Fig. \ref{fig:scheme}B), which leads to a very efficient optical pumping, since Pauli blocking is bypassed. Regime III is characterized by the dominant resonance of $\Gamma$-phonons inducing the LL transition $l=+1 \rightarrow -3$ (and $l=+3 \rightarrow -1$), which closes the laser cycle and therefore enables an efficient laser operation. In regime II, both supporting channels are resonant, however, the counteracting processes (sketched in red in Fig. \ref{fig:scheme}B) are also important (red filled curve). At about $\unit[2]{T}$ and $\unit[3]{T}$ they give rise to a very fast decay of $\rho_2$. As a consequence, the pumped inversion $\Delta_{\text{pump}}$ can not exceed $\Delta_{\rm{th}}$ and no lasing occurs. 

To sum up, at the chosen pump intensity the laser frequency can be externally tuned by applying magnetic fields in three zones between $0.5$ and $\unit[5]{T}$, where favorable phonon resonances occur.

\begin{figure}[!t]
\begin{centering}
\includegraphics[width=0.85\columnwidth]{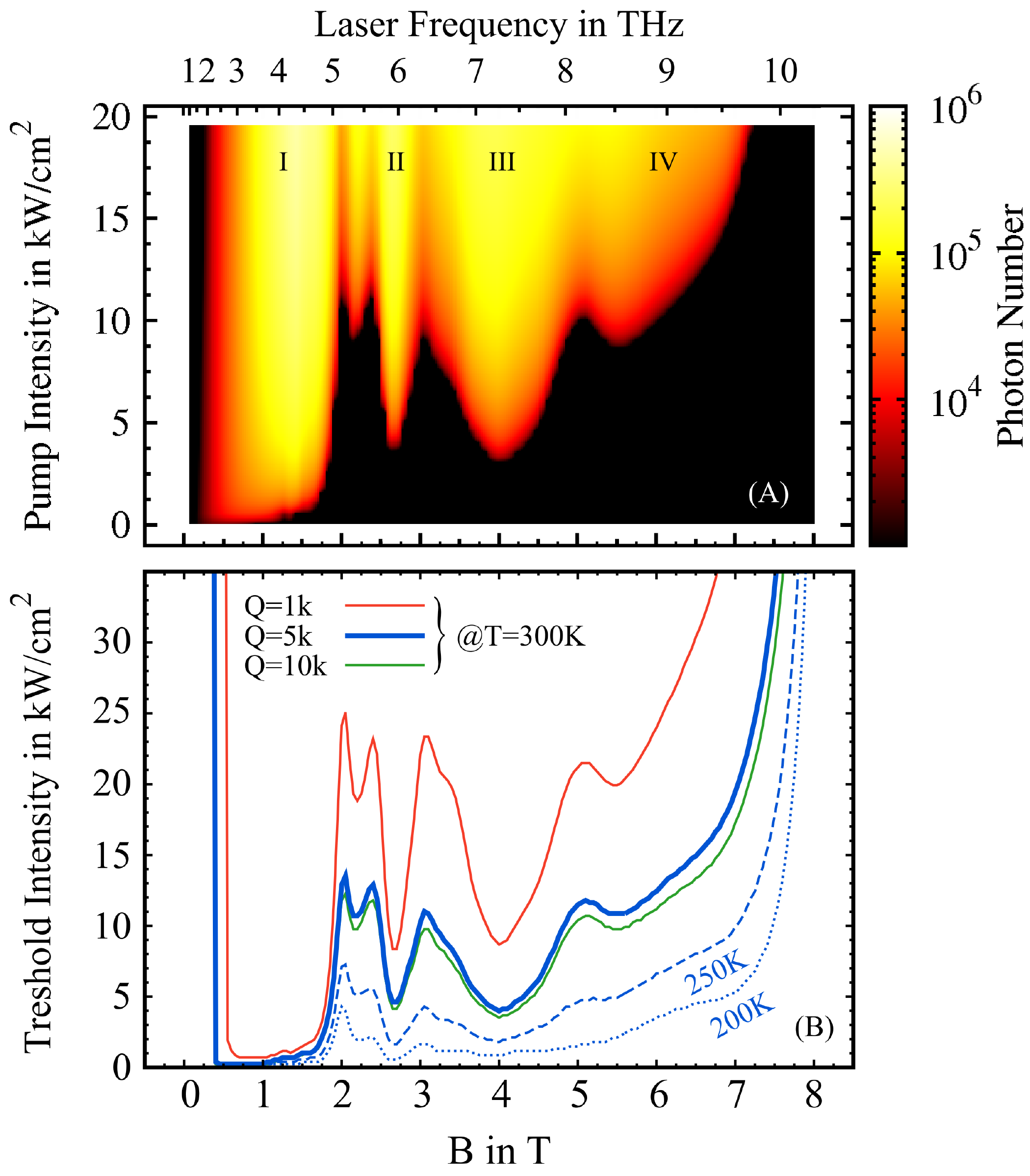}
\par\end{centering}
\caption{\textbf{Laser threshold.} 
(A) Photon number in quasi-equilibrium as a function of magnetic field and pump intensity. For each magnetic field there is a minimum pump 
intensity to achieve a sufficient population inversion to induce lasing. The magnetic field dependence of the laser threshold and the photon generation efficiency above threshold reflect the 
interplay of laser cycle supporting and counter acting phonon resonances.
(B) Magnetic field dependence of the threshold pump intensity for different cavity quality factors Q and temperatures T (with Q=5000). High qualities and low temperatures act in favor of a lower laser threshold. }
\label{fig:pump} 
\end{figure}

\textbf{Optimal conditions.} 
So far, we have fixed the pump intensity, the cavity quality factor, and the temperature. Here, we vary these experimentally accessible quantities aiming at optimal conditions for lasing. 
Figure \ref{fig:pump}A illustrates the number of photons within quasi-equilibrium as a function of the pump intensity and the magnetic field. 
Within the black areas, the pump intensity is too low to establish lasing. The pronounced line between dark and bright areas denotes the threshold intensity as a function of the magnetic field. 
Above the threshold intensity, $\Delta_{\text{pump}}$ exceeds  $\Delta_{\rm{th}}$ and the emission of coherent light occurs. The general upward trend of the threshold intensity is owed to the decrease of 
the pump transition rate with the magnetic field. Moreover, the peaks within the threshold curve are caused by the phonon resonances counteracting the population inversion  and are equivalent to the peaks of the red filled curve in Fig. \ref{fig:nonrad}. We observe that also for $B>\unit[5]{T}$ lasing can occur at high pump intensities of above $\unit[10]{kW/cm^2}$.  For very high magnetic fields with $B>\unit[7]{T}$, the threshold intensity strongly increases, since the pump efficiency decreases and a new phonon-induced counteracting relaxation process with 
 $l=+2 \rightarrow 0 \rightarrow -2$ becomes resonant.  This investigation shows that - provided sufficient pump power - the proposed laser design is in principle tunable over a broad spectral range. At a pump intensity of $\unit[20]{kW/cm^2}$,  the Landau-level laser can be continuously tuned in a range  $3-\unit[9.5]{THz}$ by applying magnetic fields of $\unit[0.7-7]{T}$.

Furthermore, the laser threshold can be lowered by improving the experimental conditions. Figure \ref{fig:pump}B shows the magnetic field dependence of the threshold pump intensity for different
cavity quality factors $Q$ and different temperatures $T$. In general, high quality factors and low temperatures lead to an overall decrease of the laser threshold. 
The influence of the Q factor can be explained by the dependence $\Delta_{\rm{th}} \propto \kappa^2 \propto Q^{-2}$, cf.  Eq. \ref{eq:/Delta_th}. That means, the higher the photon lifetime, the lower the minimum gain to compensate cavity losses. However, for $Q \longrightarrow \infty$ the minimum pump intensity still has to be sufficient to invert the inital thermal LL occupation resulting in a saturation behavior for higher $Q$ values. Thus, cooling the system has a much higher impact on the threshold, cf. the dashed lines in Fig. \ref{fig:pump}B. 

In conclusion, we predict a strategy to achieve coherent terahertz laser emission exploiting the unique properties of graphene in magnetic fields. Based on a microscopic and fully quantum-mechanical study of the coupled electron, phonon, and photon dynamics in optically pumped Landau-quantized graphene coupled to an optical cavity, we show that the emission of coherent terahertz radiation can be obtained under feasible experimental conditions. Provided an adequate cavity and sufficient pump power, the laser frequency can be externally tuned in the range of $3-\unit[9.5]{THz}$ by applying magnetic fields of $\unit[0.7-7]{T}$. The presented work provides a concrete recipe for the experimental realization of tunable graphene-based terahertz laser systems.

\section*{Acknowledgments}
We acknowledge financial support from the EU Graphene Flagship, the Swedish Research Council (VR),
 and the Deutsche Forschungsgemeinschaft (DFG) through SPP 1459. Furthermore, we thank A. Knorr (TU Berlin) and S. Winnerl (Helmholtz-Zentrum Dresden-Rossendorf) for inspiring discussions on Landau level lasers.

\newpage
\onecolumngrid
\appendix

\noindent \begin{center}
\textbf{\large Proposal for a tunable graphene-based terahertz Landau-level laser   \\ 
\vspace*{5mm}
--SUPPLEMENTARY MATERIAL--}
\par\end{center}

\begin{center}
Samuel Brem$^{1*}$, Florian Wendler$^{1}$ and Ermin Malic$^{2}${\small }\\
{\small $^{1}$ }\textit{\small Institute of Theoretical Physics,
Nonlinear Optics and Quantum Electronics, }\\
\textit{\small Technical University Berlin, Hardenbergstrasse 36,
Berlin 10623, Germany.}{\small }\\
{\small $^{2}$ }\textit{\small Department of Applied Physics, Chalmers
University of Technology, SE-412 96 G\"oteborg, Sweden}
\par\end{center}{\small \par}

\section*{Many-particle Hamilton operator}

\noindent The temporal evolution of electrons in Landau-quantized graphene coupled to a set of photon and phonon modes is determined by the many-particle Hamilton operator
\begin{equation} 
H=H_\text{el}+H_\text{ph}+H_\text{pt}\label{H_gesamt}\\.
\end{equation}
The electronic part reads
\begin{eqnarray}
\label{H_el}
H_\text{el}  \hspace{-3pt}=\hspace{-3pt}  H_\text{0,el}+H_\text{el-el}+H_\text{el-l}
  =  \sum_{i} \varepsilon_i a^{\dagger}_i a_i + \dfrac{1}{2}\sum_{ijkl} V_{ijkl}  a^{\dagger}_i a^{\dagger}_j a_k a_l -i\hbar 
\dfrac{e_0}{m_0} \sum_{ij} \mathbf{M}_{ij}\cdot \mathbf{A}(t) a^{\dagger}_i a_j,
\end{eqnarray}
and is constituted by the electronic creation and annihilation operators $a^{\dagger}_i$ and $a_i$. Here the compound index $i=(l_i,m_i,s_i,\xi_i)$ determines the electronic state\cite{Goerbig2011Electronic,wendler2015ultrafast}, containing the Landau level index $l=...,-2,-1,0,1,2, ...$, the quantum number $m=0,1,..,N_B-1$, 
which can be associated with the position of the cyclotron orbits in the graphene plane of surface $A$  ($N_B = A e_0 B/(2 \pi \hbar)$ is the number of magnetic flux quanta within the graphene plane), the spin $s=\pm 1/2$ and valley index $\xi=\pm 1$.
We include the free contribution of carriers with eigenenergies $\varepsilon_i$ (cf. the manuscript), the carrier-carrier interaction determined 
by the Coulomb matrix element $ V_{ijkl}$, and a semi-classical carrier-light coupling, which is given by the optical matrix element $\mathbf{M}_{ij}=\langle i \vert \nabla \vert j \rangle$ and the local vector potential $\mathbf{A}(t)$. 
The elementary charge and the electron mass  are denoted by $e_{0}$ and $m_0$, respectively. The tight-binding expressions of the electronic eigenenergies, eigenfunctions and all matrix elements can be found in our review article about Landau-quantized graphene\cite{wendler2015ultrafast}.
The semi-classical carrier-light coupling is used to describe the interaction with the optical pump field, whereas the light of the laser mode is treated fully quantum mechanically.

The phonon (photon) part of the Hamiltonian denoted with the subscript 'ph' ('pt') is given by
\begin{eqnarray}
\label{H_ph}
H_\text{ph}  &=&  H_\text{0,ph}+H_\text{el-ph}
  =  \sum_{\nu \mathbf{q}} \hbar \Omega_{\nu \mathbf{q}} b^{\dagger}_{\nu \mathbf{q}} b_{\nu \mathbf{q}} + \sum_{i j \nu 
\mathbf{q}} G_{ij}^{\nu \mathbf{q}} a^{\dagger}_i a_j (b_{\mathbf{q}\nu}+b^{\dagger}_{-\mathbf{q}\nu})\\
\label{H_pt}
H_\text{pt} & = & H_\text{0,pt}+H_\text{el-pt}
  =  \sum_{\mu} \hbar \omega_{\mu} c^{\dagger}_{\mu} c_{\mu} -i\hbar \sum_{ij\mu} ( g^{\mu}_{ij} 
a^{\dagger}_i a_j c_{\mu} - g^{\mu \ast}_{ij} a^{\dagger}_j a_i c^{\dagger}_{\mu} ) 
\end{eqnarray}
and includes phononic (photonic) creation operators $b^{\dagger}_{\nu \mathbf{q}}$ ($c^{\dagger}_{\mu}$) corresponding to 
the mode $\nu$ ($\mu$) and the phonon momentum $\mathbf{q}$.  It consists of a free part given by the phonon (photon) frequency  $\Omega_{\nu \mathbf{q}}$ ($\omega_{\mu}$) and an interaction part including the carrier-phonon (-photon) matrix element $ G_{ij}^{\nu \mathbf{q}}$ ( $g^{\mu}_{ij}$).

The electron-photon Hamiltonian can be deduced from the semi-classical electron-light coupling by quantizing the 
vector potential $\mathbf{A}$ and expanding it in normal modes.
Hence, the electron-photon matrix element is given by
\begin{equation}
 g_{ij}^{\mu}=\dfrac{e_0}{m_0} \sqrt{\dfrac{\hbar}{2\epsilon_0  V \omega_\mu}} \mathbf{M}_{ij} \cdot \mathbf{e}_\mu,
\end{equation}
with the normalized polarization vector of the photon mode $\mathbf{e}_\mu$
and the quantization volume $V$, which in case of a laser is equal to the volume of the cavity.

\bigskip{}

\section*{Equations of Motion}

\noindent We evaluate the Heisenberg equation of motion  $i\hbar \partial_t \langle \mathcal{O}\rangle=\langle[\mathcal{O},H]\rangle$ to determine the temporal evolution of the occupation probabilities of electronic eigenstates 
$\rho_{i}=\langle a^{\dagger}_i a_i \rangle$ and the photon numbers $n_{\mu}=\langle c^{\dagger}_{\mu} c_{\mu} \rangle$.
To prove whether coherent laser light is emitted from graphene, we also track the temporal evolution of the photon statistics via the second-order correlation
function  $g^{(2)}$, which for zero delay time is given by 
\begin{equation} \label{eq:g2}
 g^{(2)}_{\mu}(t)= \dfrac{\langle c^{\dagger}_{\mu} c^{\dagger}_{\mu} c^{\phantom\dagger}_{\mu} c^{\phantom\dagger}_{\mu} \rangle(t)}{\langle c^{\dagger}_{\mu} c^{\phantom\dagger}_{\mu} \rangle (t)^2}= 2+ \dfrac{h_\mu(t)}{n_\mu(t)^2}.
\end{equation}
Coherent laser light (Poisson statistics) is characterized by $g^{(2)}(t)=1$, whereas $g^{(2)}(t)>1$ holds for thermal and $g^{(2)}(t)<1$ for non-classical light \cite{scully1997quantum}. To calculate $g^{(2)}$ we need to consider the evolution of the photon-photon correlation 
$h_{\mu}(t)=\langle c^{\dagger}_{\mu} c^{\dagger}_{\mu} c^{\phantom\dagger}_{\mu} c^{\phantom\dagger}_{\mu} \rangle^c (t)$. To this end,  we calculate all relevant
 electron-photon-correlations up to the quadruplet level\cite{Gies2007Semiconductor,Jago2015Graphene}, thus including equations for  $T_i^{\mu}(t)=\langle c^{\dagger}_{\mu} a^{\dagger}_i a^{\phantom\dagger}_{i} c^{\phantom\dagger}_{\mu}\rangle^c (t)$ and
$U^{\mu}_{ij}(t)=\langle c^{\dagger}_{\mu} c^{\dagger}_{\mu} a^{\dagger}_i a^{\phantom\dagger}_{j} c^{\phantom\dagger}_{\mu}\rangle^c(t)$. 
Carrier-carrier and carrier-phonon correlations beyond doublets are neglected. 
We obtain the following set of coupled differential equations:
 \begin{eqnarray}
\dfrac{d}{dt} \rho_{i}  & = &  2 \sum_{\mu, {j}} \Re\lbrace \lvert g_{ji}^{\mu}\lvert^2 S_{ij}^{\mu} - \lvert g_{ij}^{\mu}\lvert^2 S_{ji}^{\mu} \rbrace +\sum_{j} P_{ij} 
(\rho_{j}-\rho_i) + \Gamma^{\text{in}}_i (1-\rho_{i}) - \Gamma^{\text{out}}_i \rho_{i}\\
\dfrac{d}{dt} n_\mu  & = &  2 \sum_{ij} \lvert g_{ij}^{\mu}\lvert^2 \Re\lbrace S_{ji}^{\mu} \rbrace -2\kappa_\mu(n_\mu-n^{0}_\mu) \label{eq:nPunkt}\\
\dfrac{d}{dt} S_{ij}^{\mu}  & = &  i(\omega_{ij}+\omega_\mu+i\kappa_\mu+i\gamma_{ij}) S_{ij}^{\mu} 
 + \rho_{j} (1-\rho_i) -  n_\mu(\rho_i-\rho_{j}) -T_i^{\mu} + T_{j}^{\mu} \\
\dfrac{d}{dt} T_{i}^{\mu}  & = &  -(2\kappa_\mu+\gamma_{ii}) T_{i}^{\mu}
+ 2 \sum_{ {j}} \Re\lbrace \lvert g_{ji}^{\mu}\lvert^2 U_{ij}^{\mu} -\lvert g_{ij}^{\mu}\lvert^2 U_{ji}^{\mu}\rbrace \nonumber\\ 
  &  & +  2 \sum_{ {j}} \Re\lbrace  \lvert g_{ji}^{\mu}\lvert^2 S_{ij}^{\mu}(n_\mu+1-\rho_i) - \lvert g_{ij}^{\mu}\lvert^2 S_{ji}^{\mu} (n_\mu+\rho_i)\rbrace\\
 \dfrac{d}{dt} U_{ij}^{\mu}  & = &  i(\omega_{ij}+\omega_\mu+3i\kappa_\mu+i\gamma_{ij}) U_{ij}^{\mu} -2 \lvert g_{ji}^{\mu}\lvert^2 (S_{ij}^{\mu})^2  \nonumber \\
 &  & -  h_\mu(\rho_i-\rho_{j}) -2 n_\mu(T_i^{\mu}-T_{j}^{\mu})   + 2(1-\rho_i)T_{j}^{\mu} -2\rho_{j} T_i^{\mu} \\
\dfrac{d}{dt} h_\mu  & = &  4 \sum_{ij} \lvert g_{ij}^{\mu}\lvert^2 \Re\lbrace U_{ji}^{\mu} \rbrace - 4\kappa_\mu h_\mu \label{eq:hPunkt}, 
\end{eqnarray}
where we have rescaled $S_{ij}^{\mu}\longrightarrow g_{ji}^{\mu} S_{ij}^{\mu}$ and $U_{ij}^{\mu}\longrightarrow g_{ji}^{\mu} U_{ij}^{\mu}$ for simplicity.
Further, $\omega_{ij}=(\varepsilon_i-\varepsilon_j)/\hbar$ denotes the electronic transition frequency and the finite photon lifetime $(2\kappa_\mu)^{-1}=Q/\omega_\mu$ accounts for cavity losses, which are determined
by the cavity quality factor $Q$.  
The Coulomb and phonon interactions are treated within second order Born-Markow approximation\cite{malic2013graphene}, which gives rise 
to the scattering rates $\Gamma^\text{in/out}_i=\Gamma^\text{in/out,el}_i+\Gamma^\text{in/out,ph}_i$ with,
\begin{eqnarray} \label{ScattRates}
\Gamma^\text{in,el}_i  & = &  \dfrac{2 \pi}{\hbar^2} \sum_{abc} V_{abci}(V_{ciab}-V_{icab}) \rho_a \rho_b 
(1-\rho_c) \mathcal{L} (\gamma_{ac}+\gamma_{bi}, \omega_{ac}+ \omega_{bi}) \\
\Gamma^\text{out,el}_i  & = &  \dfrac{2 \pi}{\hbar^2} \sum_{abc} V_{abci}(V_{ciab}-V_{icab})(1- \rho_a) 
(1-\rho_b) \rho_c \mathcal{L} (\gamma_{ac} +\gamma_{bi}, \omega_{ac}+ \omega_{bi}) \\
\Gamma^\text{in,ph}_i  & = &  \dfrac{2 \pi}{\hbar^2} \sum_{j\nu\mathbf{q}} \vert G_{ij}^{\nu\mathbf{q}} 
\vert^2 \rho_j \Big( N_{\nu \mathbf{q}} \mathcal{L} (\gamma_{ij}, \omega_{ji}+ \Omega_{\nu\mathbf{q}})  + 
(N_{\nu \mathbf{q}} +1) \mathcal{L} ( \gamma_{ij}, \omega_{ji}- \Omega_{\nu\mathbf{q}}) \Big) \\
\Gamma^\text{out,ph}_i  & = &  \dfrac{2 \pi}{\hbar^2} \sum_{j\nu\mathbf{q}} \vert G_{ij}^{\nu\mathbf{q}} 
\vert^2 (1-\rho_j) \Big( N_{\nu \mathbf{q}} \mathcal{L} (\gamma_{ij}, \omega_{ij}+ \Omega_{\nu\mathbf{q}})  
+ (N_{\nu \mathbf{q}} +1) \mathcal{L} (\gamma_{ij}, \omega_{ij}- \Omega_{\nu\mathbf{q}}) \Big).
\end{eqnarray}
Applying a bath approximation, the phonon number $N_{\nu \mathbf{q}}=\langle b^{\dagger}_{\nu \mathbf{q}} b_{\nu \mathbf{q}}\rangle$ can be fixed to the thermal occupation (Bose-Einstein statistics). This is a good approximation in the considered laser regime.
Phonon scattering is considered only for the dominant optical phonon modes $\Gamma \text{TO}$, $\Gamma \text{LO}$,
$\text{KTO}$ and $\text{KLO}$ \cite{piscanec2004kohn,Maultzsch2004Phonon}, with $\epsilon_{\text{KLO}}=\unit[151]{meV}$, $\epsilon_{\text{KTO}}=\unit[162]{meV}$, $\epsilon_{\Gamma \text{LO}}=\unit[198]{meV}$ and $\epsilon_{\Gamma \text{TO}}=\unit[192]{meV}$ (Einstein approximation).
 
The energy conservation is softened due to the Lorentzian broadening
\begin{equation} \label{Lorentz}
 \mathcal{L}(\gamma, \omega)= \frac{1}{\pi} \dfrac{\gamma}{\gamma^2+\omega^2},
\end{equation}
whose width is given by the dephasing $\gamma_{ij}$, which is self-consistently determined \cite{malic2013graphene} considering
impurity and many-particle scattering. It reads
\begin{eqnarray} \label{Broadening}
 \gamma_{ij} & = & \gamma_{\text{imp}}+\gamma^\text{el}_{ij}+\gamma^\text{ph}_{ij}\quad \text{with} \quad \gamma^\text{el/ph}_{ij}  =  \dfrac{1}{2}\sum_{k=i,j}(\Gamma^\text{in,el/ph}_k + \Gamma^\text{out,el/ph}_k) .
\end{eqnarray}
Since the scattering rates $\Gamma_i$ themselves depend on the dephasing, they are determined iteratively starting with $\gamma_{ij}=\gamma_{\text{imp}}$.

The disorder contribution to the equation of motion is derived within a selfconsistent Born approximation, following the approach of Shon and Ando \cite{ando1974theory, shon1998quantum}.
We assume \cite{wendler2015ultrafast},
\begin{equation}
 \gamma_\text{imp}=\dfrac{v_F}{l_B \sqrt{A_\text{imp}}}=v_F \sqrt{\dfrac{e_0 B}{\hbar A_\text{imp}}},
\end{equation}
where $A_\text{imp}$ denotes a dimensionless parameter characterising the scattering strenght of the impurity
potential\cite{ando1974theory,shon1998quantum}. Since this parameter is not accessible in experiments, we
assume the impurity parameter $A_{\text{imp}}=\unit[420]{}$, since the corresponding broadening of $\unit[2.5] {meV}$ at $B=\unit[4]{T}$
is in good agreement with experimental studies of linewidths in absorption spectra \cite{mittendorff2015carrier}. 

To obtain the optical pump rate $P_{ij}$, the equation of motion for the microscopic polarization 
$p_{ij}=\langle a^{\dagger}_i a_j \rangle$ is solved within the Markow and rotating wave approximation. 
For a constant optical pump field with the frequency $\omega_\text{P}$, intensity $I_\text{P}$, and polarization $\mathbf{e}_\text{P}$ one obtains:
\begin{equation} \label{Pumprate}
 P_{ij}=\Big(\dfrac{e_0}{m_0}\Big )^2  \dfrac{\pi I_\text{P}}{\epsilon_0 c \omega_\text{P}^2} \vert \mathbf{M}_{ij} \cdot  
\mathbf{e}_\text{P}\vert^2 \Big( \mathcal{L}(\gamma_{ij}, \omega_{ij}+\omega_\text{P}) + \mathcal{L}(\gamma_{ij},\omega_{ij}-\omega_\text{P}) \Big ) 
\end{equation} 

The degeneracy of Landau levels in spin $s=\pm 1/2$, valley $\xi=\pm 1$ and quantum number $m=0,1,..,N_B-1$ gives rise to a total amount of $4N_B$ LLs with the same energy.
Our numerical calculations show that for $N_B \gg 1$ the electronic dynamics only depend on the Landau level index $l$, i.e. all degenerated levels behave equally. 
 Thus, we define averaged quantities,
\begin{eqnarray} \label{Average}
\rho_l & = & \dfrac{1}{4N_B} \sum_{m,s,\xi} \rho_{(l,m,s,\xi)} \\[5pt]
S_{l,l'}^{\mu} & = & \dfrac{1}{4 N_B} \sum_{m,s,\xi} S^{\mu}_{(l,m,s,\xi)(l',m,s,\xi)},
\end{eqnarray} 
where we only have to consider $s$-,$\xi$- and $m$-diagonal polarizations, since other polarizations are forbidden by selection rules \cite{wendler2015ultrafast}.
$T_i^{\mu}$ and $U_{ij}^{\mu}$ are treated in analogous manner. As we assume that all observables are in good approximation independent of $m,s$ and $\xi$, we set $\rho_l \approx \rho_{(l,m,s,\xi)}$, $S_{l,l'}^{\mu} \approx S^{\mu}_{(l,m,s,\xi)(l',m,s,\xi)}$ and so forth.
Hence, the photon generation rate in Eq. \ref{eq:nPunkt} can be written as,

\begin{eqnarray} \label{Degeneracy}
\sum_{ij} \lvert g_{ij}^{\mu}\lvert^2 S_{ji}^{\mu} \approx 4 N_B \sum_{l_i, l_j} \lvert g_{l_il_j}^{\mu}\lvert^2 S_{l_jl_i}^{\mu},
\end{eqnarray}

where\cite{wendler2015ultrafast} $g_{ij}^{\mu}=g_{l_il_j}^{\mu} \delta_{m_i,m_j}\delta_{s_i,s_j}\delta_{\xi_i,\xi_j}$.
The same procedure applies for the sums in Eq. \ref{eq:hPunkt}.

\end{document}